\documentclass[conference, 10pt]{IEEEtran}
\usepackage{graphics}
\usepackage{amsmath,amsfonts,amssymb,varioref,mathrsfs,verbatim}
\usepackage{epsfig}
\usepackage{array}
\usepackage[english]{babel}
\usepackage{color}
\usepackage{babel}

\usepackage{color}
\renewcommand{\figurename}{Fig.}
\renewcommand{\figurename}{Fig.}
\addto\captionsenglish{\renewcommand{\figurename}{Fig.}}

\begin{document}
\pagenumbering{gobble}

\title{Exploiting Multi-Path for Safeguarding \\ mmWave Communications Against Randomly Located Eavesdroppers}

\author{Rohith Talwar, Nancy Amala, George Medina, Akshadeep Singh Jida, and Mohammed E. Eltayeb \\Department of Electrical \& Electronic Engineering  \\ 
California State University, Sacramento, USA \\ Emails: \{rohithtalwar, nancyamalajosephraj, gm739, asjida, mohammed.eltayeb\}@csus.edu }
\maketitle

\begin{abstract}
Communication in the millimeter-wave (mmWave) band has recently been proposed to enable giga-bit-per-second data rates for next generation wireless systems. Physical layer security techniques have emerged as a simple and yet effective way to safeguard these systems against eavesdropping attacks.  These techniques make use of the large antenna arrays available in mmWave systems to provide an array gain at the target receiver and degrade the signal quality at the eavesdropper. Despite their effectiveness, majority of these techniques are based on line-of-sight communication links between the transmitter and the receiver, and may fail in the presence of blockages or non-line-of-sight links. This paper builds upon previous work and extends physical layer security to the non-line-of-sight communication case and randomly located eavesdroppers. Specifically, the large dimensional antenna arrays in mmWave systems and the intrinsic characteristics of wireless channel are exploited to induce noise-like signals that jam eavesdroppers with sensitive receivers. Numerical results show that the proposed techniques provide higher secrecy rate when compared to conventional array and physical layer techniques based on line-of-sight links.

\end{abstract}

\section{Introduction}
The abundance of bandwidth in the millimeter-wave (mmWave) band enables high-speed,  low latency, communication to support next generation systems \cite{mag}-\cite{I3}. These systems rely on directional communication with large antenna arrays to achieve sufficient link budget \cite{ahmed}, \cite{m1}. Securing these systems against eavesdropping and privacy attacks is one major challenge \cite{sec1}-\cite{sec3}.

Security in wireless communications can be mainly grouped into cryptographic and physical layer security (PLS) strategies, where the previous incorporates key-based approaches normally applied at higher system layers. The limited transmit RF-chains, higher path-loss, and directional communication requirement of mmWave systems, however, makes the implementation of classical cryptographic techniques inefficient for these systems \cite{ahmed}, \cite{sec3}, \cite{A2}. For this reason, physical layer security has emerged as a simple but yet effective way to secure these systems and complement key-based cryptographic techniques \cite{sec1},\cite{sec3}, \cite{haji}. Further, the implementation of PLS in mmWave is considered a productive strategy  as it can result in higher communication speed, narrow communication beam and shorter transmission separation \cite{sun2018}. Physical layer security techniques utilize the large antenna arrays to enhance the communication signal (via beamforming) at the target receiver and degrade the signal quality at the eavesdropper. Directional beamforming, however, relies on line-of-sight (LoS) communication, and assumes the eavesdropper is not along the direction of communication.  In practice, the passive eavesdropper could be located in the direction of communication, and, in the case of non-light-of-sight (NLoS) communication, scatters can enable the eavesdropper to intercept the communication signal via multipath, hence rendering PHY layer security techniques ineffective. Therefore, it is critical to develop PHY layer security techniques that preserve privacy in both the LoS and NLoS cases.  

Several strategies have been adopted in the literature to enhance the secrecy of mmWave wireless systems. Switched array based schemes \cite{sec3}, \cite{sec4}, \cite{sec5}, \cite{jing2018} employ random antenna subsets  to induce artificial noise that jams potential eavesdroppers in non-receiver directions. In addition to this, the work in \cite{sec3} proposed the use of prior information obtained from sensors to opportunistically inject noise in potential eavesdropper locations.  In \cite{jung2018} and \cite{jung2019}, low-complexity PLS techniques based on analog cooperative beamforming are proposed. These techniques enhanced the secrecy rate as distributed antenna nodes result in narrower beamwidth and increased artificial noise. The work in \cite{jung2020} built upon the techniques in \cite{jung2018} and \cite{jung2019} and proposed a beamforming solution with null steering. This solution is shown to improve the secrecy performance  when the eavesdropper is in close proximity to the target receiver. Despite their capability in enhancing the transmission secrecy, those techniques require a LoS link to the target receiver and fail if the eavesdropper and the target receiver paths overlap.  For instance, path overlap can occur if the eavesdropper is located along the transmission direction of the receiver or it is in close proximity to the receiver. Scatters close to the transmitter can cause blockages and lead to NLoS links. Overlapped common channel paths (as a result of NLoS links) between the receiver and the eavesdropper increase the risk of information leakage to the eavesdropper. To safeguard mmWave systems with possible overlapped common channel paths with the eavesdropper, the work in \cite{sec6},  \cite{jing2020} , \cite{Wang} and \cite{Ramadan} considered the problem of  aligned transmitter, eavesdropper, and target receiver in the mainlobe path, i.e. mainlobe security. In \cite{sec6},  a dual beam transmission technique that ensures the mainlobe is coherent only at the target receiver's location is proposed. The work in \cite{jing2020} proposed a dynamic rotated angular beamforming technique which uses a set of frequency offset modulator to generate angle and range based transmission.  The work in \cite{Ramadan} utilized channel knowledge of eavesdroppers to enhance the secrecy rate of the target receiver. It should be noted that eavesdroppers in general are passive and their presence is unknown. The work in  \cite{Wang} proposed the use of maximum ratio combining and artificial noise injection in the null space of the receiver's channel to safeguard transmission. Maximum ratio combining, however, only increases the SNR at the target receiver, but does not safeguard the communication link against an eavesdropper with a sensitive receiver. Moreover, noise injection requires complex antenna architectures which might be impractical to implement in mmWave systems.

In this paper, we address the problem of overlapped communication channel paths between the receiver and an arbitrary eavesdropper and  propose two transmission techniques that enhance the security of mmWave systems with NLoS channels. Unlike \cite{sec6},  \cite{jing2020} , \cite{Wang} and \cite{Ramadan}, the proposed techniques do not necessarily require a LoS channel path between the transmitter and the receiver,  and can be applied using  a simple analog antenna architecture with a single RF chain at the transmitter. The first technique enhances secrecy by employing a path hopping technique, while the second technique uses random antenna subsets to beamform along multiple transmission paths. The proposed techniques facilitate secure transmission to the receiver without the need for complex antenna architectures and reduce the likelihood of information leakage to potential eavesdropper with common receiver communication paths.

\section{System Model}
We consider a mmWave system where the transmitter communicates with a receiver via NLoS communications paths $L>1$ in the presence of an eavesdropper. The transmitter consists of one RF chain and $N$ antennas and the receiver is equipped with  $N_\text{R}$ antennas. To transmit the $k$th information symbol $s(k)$, where $E[|s(k)|^2] =1$, to the receiver, the transmitter multiplies $s(k)$ by unit norm transmit vector  $\mathbf{f}=[f_1, f_2, f_3, \dots f_N] \in \mathcal{C}^N$, where $f_n$ denotes the complex weight (or phase-shift) associated with the $n$th transmitting antenna. At the receiver, the received signals on all receive antennas are combined using a combining vector $\mathbf{w}\in \mathcal{C}^{N_\text{R}}$. Assuming a narrowband block fading channel, the received signal at the receiver is given by
\begin{eqnarray}\label{rs}
y(k) =  \mathbf{w}^*\mathbf{H}\mathbf{f} s(k) + z(k), 
\end{eqnarray}
where $\mathbf{H}$ represents the mmWave channel matrix between the transmitter and the receiver, and $z(k) \sim \mathcal{CN}(0,\sigma^2)$ is the additive noise at the receiver. Adopting a geometric channel model with $L$ scatters, where each scatter is assumed to contribute to a single channel path between the transmitter and the receiver, the channel $\mathbf{H}$ is given by \cite{ahmed}
\begin{eqnarray}\label{ch1}
\mathbf{H}=\sqrt{\frac {N N_\text{R}}{L}} \sum_{l=1}^{L} \alpha_l \mathbf{a}_\text{R} (\phi_l) \mathbf{a}^*_\text{T} (\theta_l)
\end{eqnarray}
where $\phi_l$ is angle-of-arrival (AoA) at the receiver associated with the $l$th angle-of-departure (AoD) $\theta_l$ from the transmitter, $\alpha_l$ is the complex gain of the $l$th path and $\mathbf{a}_\text{R} (\phi_l)$ and  $\mathbf{a}_\text{T} (\theta_l)$ represent the receiver's and transmitter's array response defined in \cite{ahmed}. For ease of exposition, we assume a uniform linear array at the transmitter and a single antenna at the receiver. Hence, the 
the channel in (\ref{ch1}) becomes
\begin{eqnarray}\label{ch}
\mathbf{h}=\sqrt{\frac {N}{L}} \sum_{l=1}^{L} \alpha_l  \mathbf{a}_\text{T} (\theta_l),
\end{eqnarray}
where $\mathbf{a}_\text{T} (\theta_l) = \frac{1}{\sqrt{N}}[e^{-j(\frac{N-1}{2} )2\pi \frac{d}{\lambda} \cos(\theta_l)}, ..., e^{j(\frac{N-1}{2} )2\pi \frac{d}{\lambda} \cos(\theta_l)}  ]$ and the signal  in (\ref{rs}) simplifies to
\begin{eqnarray}\label{rs2}
y(k) =  \mathbf{h}^*\mathbf{f} s(k) + z(k).
\end{eqnarray}

\section{Millimeter Wave Secure Transmission}
In this section, we introduce simple transmission techniques suitable for mmWave systems with analog antenna architecture (single RF chain). Nonetheless, the proposed techniques can also be applied in systems with advanced antenna architectures. The key idea is to exploit the intrinsic randomness of the wireless channel to distort transmission to eavesdroppers, with possible overlapping channels with the receiver, without the need for a fully digital array or prior channel information of eavesdroppers.  With the assumption of local scatters at both the transmitter and receiver, the receiver's channel becomes unique at the transmitter, i.e. function of its location and scatters. This property has been exploited in finger-printing and localization techniques, see example \cite{F1}-\cite{F3} and references therein, to uniquely identify the location of the receiver. In this paper, with the assumption of full channel knowledge at the both the transmitter and the receiver, we exploit this property to jam an eavesdropper with an overlapping communication channel, i.e. eavesdropper shares a communication path with the target receiver. Instead of transmitting data symbols using the strongest channel path, the proposed transmission techniques select a random set of paths to transmit each data symbol (or data packet) to the target receiver. With the assumption that the receiver knows a-priori the selected transmission paths (this could be preprogrammed in the system) the receiver receives coherent signal,  whereas, the eavesdropper receives a noise-like signal. This noise-like signal, which we term artificial noise,  is a result of the randomized beam pattern observed at the eavesdropper. In the following sections, we further elaborate on those transmission techniques.

\subsection{Enhancing secrecy with random path selection}
In this technique, the transmitter transmits each data symbol along a random path (or AoD). Specifically, let $\theta_l \in \Phi$ be the $l$th AoD towards the target receiver and the set $\Phi$ represents all possible AoDs towards the target receiver. The transmitter's inter-antenna phase shifts are set as \cite{sec5}
  \begin{equation}\label{efbp}
\Upsilon_{n}(k) =  { \left(\frac{N-1}{2}-n \right)2\pi \frac{d}{\lambda} \cos(\theta_l)}
\end{equation}
where $\theta_l $ represents the $l$th selected AoD for the $k$th data symbol.  Based on this, the $n$th entry of the beamforming vector $\mathbf{f}(k)$ (i.e. $n$th antenna weight) becomes $f_n(k)=\frac{1}{\sqrt{N_\text{}}}e^{ j \Upsilon_{n}(k)}$. 

Using (\ref{rs2}), the received signal at the receiver along AoD $\theta_l$ becomes
\begin{eqnarray}\label{yra88a}
\begin{aligned} 
 & y_{\text{R}}(k,\theta_l)  =   \mathbf{h}^* \mathbf{f}(k) s(k) +z_\text{R} (k) \\   & =  \sqrt{\frac {N}{L}}\alpha_l \mathbf{a}^*_\text{T} (\theta_l) \mathbf{f}(k) s(k) +z_\text{R} (k) = {\frac{1}{\sqrt{L N}}} \alpha_l s(k) \times  \\
  &  \bigg( \sum_{n=0}^{N-1} e^{-j\left(\frac{N-1}{2}-n\right)  \frac{2\pi d}{\lambda}  \cos\theta_l  }  e^{j \left(\frac{N-1}{2}-n\right)  \frac{2\pi d}{\lambda}  \cos\theta_l } \bigg)  + z_\text{R} (k), \\
  &=   \underbrace{s(k)}_{\substack{\text{ information }\\\text{symbol}}}  \underbrace{\alpha_l \sqrt{\frac {N}{L}}}_{\substack{\text{effective channel}\\\text{and array gain}}}   +  \underbrace{z_\text{R}(k)}_{\substack{\text{additive }\\\text{noise}}}.
\end{aligned}
\end{eqnarray}        
where $\alpha_l$ is the $l$th path gain  gain at the receiver,  and $z_\text{R}(k)\sim\mathcal{CN}(0,\sigma^2_\text{R})$ is the additive noise at the receiver.

Similarly, the received signal at an eavesdropper along an AoD $\theta_\text{E}$  is given by
\begin{eqnarray}\label{arnl}
\begin{aligned} 
& y_{\text{E}}(k,\theta_\text{E})  =   \mathbf{h}^*\mathbf{f}(k)s(k) +z_\text{E} (k) \\& =  \sqrt{\frac {N}{L}}\alpha_l \mathbf{a}^*_\text{T} (\theta_\text{E}) \mathbf{f}(k) s(k) +z_\text{E} (k) = {\frac{1}{\sqrt{L N}}} \alpha_\text{E}s(k) \times  \\
 &  \bigg(  \sum_{n=0}^{N-1}  e^{-j\left(\frac{N-1}{2}-n\right)  \frac{2\pi d}{\lambda}  \cos\theta_\text{E} }  e^{j \left(\frac{N-1}{2}-n\right)  \frac{2\pi d}{\lambda}  \cos\theta_l }   \bigg) +\hspace{-1mm}  z_\text{E}(k)\\ 
&=  \underbrace{s(k)}_{\substack{\text{ information }\\\text{symbol}}}  \underbrace{\alpha_\text{E}}_{\substack{\text{channel }\\\text{gain}}} \underbrace{\sqrt{1/L N}B(\theta_l)}_{\substack{\text{artifical }\\\text{noise}}}  +  \underbrace{z_\text{E}(k)}_{\substack{\text{additive }\\\text{noise}}},  
   \end{aligned}
   \end{eqnarray}
where $\theta_l \in \Phi$, $z_\text{E}(k)\sim\mathcal{CN}(0,\sigma^2_\text{E})$ is the additive noise at the eavesdropper and  the random variable $B(\theta_l)$ (due to random selection of $\theta_l$) is
\begin{eqnarray}\label{beta}
B(\theta_l) &=&  \sum_{n=0}^{N-1} e^{-j\left(\frac{N-1}{2}-n\right)  \frac{2\pi d}{\lambda}  \cos\theta_\text{E} }  e^{j \left(\frac{N-1}{2}-n\right)  \frac{2\pi d}{\lambda}  \cos\theta_l }\\
&=& \sum_{n=0}^{N-1} e^{-j\left(\frac{N-1}{2}-n\right)  \frac{2\pi d}{\lambda}  (\cos(\theta_\text{E}) -\cos(\theta_l) )} \\
&=&  {N} \frac{\sin (N(\frac{\pi d}{\lambda} (\cos (\theta_\text{E}) - \cos (\theta_{l}))   )) } {\sin ((\frac{\pi d}{\lambda} (\cos (\theta_\text{E}) - \cos (\theta_{l}))   )) }.
\end{eqnarray}
Observe when the eavesdropper is located along one of the AoDs in the set $\Phi$, i.e. $\theta_\text{E} = \theta_l$, then $B(\theta_l)$ becomes deterministic with the value of ${N}$. Note this event occurs with probability $\frac{1}{L}$, where $L$ is the number of paths. Hence, it is important to have large number of paths $L$ for this technique to be efficient.

\subsection{Enhancing secrecy with joint path and antenna selection}
The secrecy performance of the random path selection technique deteriorates with limited number of scatters, i.e. low number of transmit path $L$. To improve the secrecy in a low scattering environment, we propose a joint antenna selection and path selection technique. The idea is to associate a set of $M$ antennas for the strongest transmission path and the remaining $N-M$ antennas for another randomly selected path. For each transmit symbol (or packet),  different sets of antennas are associated with the strongest path and a random secondary transmit path. Selecting the strongest path maximizes the receive signal-to-noise ratio (SNR) at the receiver. This dual antenna and path selection technique results in controlled interference in both the main and secondary transmission paths. The interference results from the contribution of the side-lobes of each transmit beam in all paths. The random antenna selection ensures that the side-lobe levels are perturbed for each transmit symbol. As the target receiver is aware of its channel, the transmit path identity, and the set of selected antennas for each path (can be preprogrammed a priori),  it is able to decode the received symbol. As the eavesdropper does not have knowledge of these parameters, it can not precancel the interference due to the perturbed side-lobes. This reduces its rate and enhances the overall secrecy rate.

Let $\mathcal{I}_M(k)$ be a random subset of $M$ antennas used to transmit the $k$th symbol along the strongest path (AoD $\theta_{\text{S}}$),  and $\mathcal{I}_L(k)$ be subset that contains the indices of the remaining antennas used to transmit the $k$th symbol along a secondary path (AoD $\theta_{{i}}$).  The $n$th antenna phase shift is set as
  \begin{equation}\label{efbp2}
\Upsilon_{n}(k)  = \left\{
               \begin{array}{ll}
     { \left(\frac{N-1}{2}-n \right)2\pi \frac{d}{\lambda} \cos(\theta_{\text{S}})}, & \hbox{  $n\in \mathcal{I}_M(k)$  } \\
                   { \left(\frac{N-1}{2}-n \right)2\pi \frac{d}{\lambda} \cos(\theta_i)}, & \hbox{  $n\in \mathcal{I}_L(k)$  }  \\
               \end{array}
               \right.
\end{equation}
and the received signal at the target receiver becomes
\begin{eqnarray}\nonumber
\begin{aligned} \label{yra8}
 & \hspace{-0mm} y_{\text{R}}(k,\theta_{\text{S}},\theta_i)  =  \mathbf{h}^*\mathbf{f}(k)s(k) +z_\text{R} (k) =  {\frac{1}{\sqrt{L N_\text{}}}} s(k) \times  \\
  &  \bigg(  \sum_{n \in \mathcal{I}_M(k)} \alpha_\text{S}e^{-j\left(\frac{N-1}{2}-n\right)  \frac{2\pi d}{\lambda}  \cos \theta_\text{S} }  e^{j \left(\frac{N-1}{2}-n\right)  \frac{2\pi d}{\lambda}  \cos \theta_\text{S} }    \\  & \hspace{-0mm} + \hspace{-1mm} \sum_{n \in \mathcal{I}_L(k)}  \alpha_\text{S} e^{-j\left(\frac{N-1}{2}-n\right)  \frac{2\pi d}{\lambda}  \cos \theta_\text{S} }  e^{j \left(\frac{N-1}{2}-n\right)  \frac{2\pi d}{\lambda}  \cos\theta_{i} }  \\  & \hspace{-0mm} + \hspace{-1mm}    \sum_{n \in \mathcal{I}_L(k)} \alpha_i  e^{-j\left(\frac{N-1}{2}-n\right)  \frac{2\pi d}{\lambda}  \cos\theta_i }  e^{j \left(\frac{N-1}{2}-n\right)  \frac{2\pi d}{\lambda}  \cos\theta_{i} } \\  & \hspace{-0mm} + \hspace{-1mm}    \sum_{n \in \mathcal{I}_M(k)} \alpha_i e^{-j\left(\frac{N-1}{2}-n\right)  \frac{2\pi d}{\lambda}  \cos \theta_{i} }  e^{j \left(\frac{N-1}{2}-n\right)  \frac{2\pi d}{\lambda}  \cos \theta_\text{S} } \bigg)  \\  & \hspace{-0mm} + \hspace{-1mm}  z_\text{R} (k)
  \\  & \hspace{-0.5mm}  =  \underbrace{s(k)}_{\substack{\text{ information }\\\text{symbol}}} \underbrace{{\frac{1}{\sqrt{L N_\text{}}}}  \bigg( \alpha_\text{S} M + \alpha_\text{i} (N-M) + \beta_\text{R}\bigg)}_{\substack{\text{effective beamforming and }\\\text{channel gain}}} + \underbrace{z_\text{R}(k)}_{\substack{\text{additive }\\\text{noise}}}, 
\end{aligned} \\
\end{eqnarray}         
where $\beta_\text{R}$ is given by
\begin{eqnarray}\label{beta2}
\nonumber &\beta_\text{R}& \hspace{-2mm} = \hspace{-2mm} \sum_{n \in \mathcal{I}_L(k)}  \alpha_\text{S} e^{j\left(\frac{N-1}{2}-n\right)  \frac{2\pi d}{\lambda}  (\cos \theta_i-\cos \theta_\text{S})   } +  \\ & & \hspace{-2mm}  \sum_{n \in \mathcal{I}_M(k)} \alpha_i e^{j\left(\frac{N-1}{2}-n\right)  \frac{2\pi d}{\lambda} (\cos \theta_\text{S}-\cos \theta_{i})}.
\end{eqnarray}   
Note $\beta_\text{R}$ can be precalculated at the receiver since it has prior knowledge of the channel and the transmit antenna set along each path.

The eavesdropper can intercept the transmitted signal via the transmitter's side lobe or main lobe (this occurs when the eavesdropper is located along the the transmitter's main lobe). We will investigate these two scenarios separately.
Assuming the eavesdropper can only intercept communication via the side lobe, the received signal at the eavesdropper along its AoD $\theta_\text{E} \not \in \Phi$ is given by
\begin{eqnarray}\label{ye0} \nonumber
\begin{aligned} 
 & \hspace{-0mm} y_{\text{E}}(k,\theta_{\text{S}},\theta_i,\theta_{\text{E}})  =  \mathbf{h}^*\mathbf{f}(k) s(k)+z_\text{E} (k) =  {\frac{1}{\sqrt{L N_\text{}}}} s(k) \alpha_\text{E} \times  \\
  &  \bigg(  \sum_{n \in \mathcal{I}_M(k)} e^{-j\left(\frac{N-1}{2}-n\right)  \frac{2\pi d}{\lambda}  \cos \theta_\text{E} }  e^{j \left(\frac{N-1}{2}-n\right)  \frac{2\pi d}{\lambda}  \cos \theta_\text{S} }  +  \\  & \hspace{3mm} \sum_{n \in \mathcal{I}_L(k)}  e^{-j\left(\frac{N-1}{2}-n\right)  \frac{2\pi d}{\lambda}  \cos \theta_\text{E} }  e^{j \left(\frac{N-1}{2}-n\right)  \frac{2\pi d}{\lambda}  \cos\theta_{i} }   \bigg)  + z_\text{E} (k)
\end{aligned} 
\end{eqnarray}     
\begin{eqnarray}\label{ye1} 
 & \hspace{-30mm}  =  \underbrace{s(k)}_{\substack{\text{ information }\\\text{symbol}}} \underbrace{   \alpha_\text{E} \beta_\text{E}}_{\substack{\text{effective artificial }\\\text{noise}}} + \underbrace{z (k)}_{\substack{\text{additive }\\\text{noise}}}, 
 \end{eqnarray}      
where the noise term $\beta_\text{E}$ is given by
\begin{eqnarray}\label{beta2}
\nonumber &\beta_\text{E}& \hspace{-2mm} =  \frac{1}{\sqrt{L N}}\sum_{n \in \mathcal{I}_L(k)}   e^{j\left(\frac{N-1}{2}-n\right)  \frac{2\pi d}{\lambda}  (\cos \theta_\text{S}-\cos \theta_\text{E})   } +  \\ & & \hspace{-2mm}  \sum_{n \in \mathcal{I}_M(k)}  e^{j\left(\frac{N-1}{2}-n\right)  \frac{2\pi d}{\lambda} (\cos \theta_i -\cos \theta_\text{E})}.
\end{eqnarray}

From (\ref{beta2}) we observe that the noise term $\beta_\text{E}$ is random since the eavesdropper is unaware of the transmit angles $\theta_i$ and $\theta_\text{s}$ and the associated transmit antenna sets $\mathcal{I}_L(k)$ and $\mathcal{I}_M(k)$. This induces artificial noise that jams the eavesdropper. We also observe from (\ref{ye1}) that high channel gain $\alpha_\text{E}$  increases the artificial noise at the eavesdropper, hence, no significant rate gain is expected with high channel gain.

Considering the second scenario, where we assume the eavesdropper can only intercept the communication link via the main lobe, the received signal at the eavesdropper along an AoD $\theta_\text{E} = \theta_\text{S}$  becomes
\begin{eqnarray}\label{ye2}
\begin{aligned} \nonumber
 &  y_{\text{E}}(k,\theta_{\text{S}},\theta_i,\theta_{\text{E}})  =  \mathbf{h}^*\mathbf{f}(k) s(k) +z_\text{E} (k)=  {\frac{1}{\sqrt{L N_\text{}}}} s(k) \alpha_\text{E} \times  \\
  &  \bigg(  \sum_{n \in \mathcal{I}_M(k)} e^{-j\left(\frac{N-1}{2}-n\right)  \frac{2\pi d}{\lambda}  \cos \theta_\text{E} }  e^{j \left(\frac{N-1}{2}-n\right)  \frac{2\pi d}{\lambda}  \cos \theta_\text{S} }  +  \\  & \hspace{3mm} \sum_{n \in \mathcal{I}_L(k)}  e^{-j\left(\frac{N-1}{2}-n\right)  \frac{2\pi d}{\lambda}  \cos \theta_\text{E} }  e^{j \left(\frac{N-1}{2}-n\right)  \frac{2\pi d}{\lambda}  \cos\theta_{i} }   \bigg)  + z_\text{E} (k)
\end{aligned}
\end{eqnarray}        
\vspace{-5mm}
\begin{eqnarray}\label{ye2}
& \hspace{-30mm}  =  \underbrace{s(k)}_{\substack{\text{ information }\\\text{symbol}}} \underbrace{  {\frac{\alpha_\text{E} \hat{\beta}_\text{E}}{\sqrt{L N_\text{}}}} }_{\substack{\text{effective artificial }\\\text{noise}}} + \underbrace{z_\text{E} (k)}_{\substack{\text{additive }\\\text{noise}}}, 
\end{eqnarray}
\vspace{-1mm}
where the noise term $\hat{\beta}_\text{E}$ is given by
\begin{equation}\label{beta3}
\hat{\beta}_\text{E}  = \left\{
               \begin{array}{ll}
               {    M+    \sum_{n \in \mathcal{I}_L(k)}  e^{-j\left(\frac{N-1}{2}-n\right)  \gamma(\theta_{i}, \theta_\text{E}) } }   , & \hbox{\hspace{-3mm} if  $\theta_\text{E} = \theta_\text{S}$  } \\
{N-M+ \sum_{n \in \mathcal{I}_M(k)}  e^{-j\left(\frac{N-1}{2}-n\right)  \gamma(\theta_\text{S}, \theta_\text{E})  } }   , & \hbox{\hspace{-3mm} if  $\theta_\text{E} = \theta_i$  }
               \end{array}
               \right.
\end{equation}
and $\gamma(\theta_x,\theta_\text{E}) = \frac{2\pi d}{\lambda}  (\cos\theta_{x}-\cos \theta_\text{E})$. From (\ref{beta3}) we observe that the noise term $\hat{\beta}_\text{E}$ consists of a constant term and a random term (function of transmit antenna sets and AoD). Hence, to maintain the randomness of the term $\hat{\beta}_\text{E}$ both transmit subsets should be sufficiently large. This maximizes the randomness of  $\hat{\beta}_\text{E}$  and, as a result, the artificial noise at the eavesdropper.

\section{Performance Evaluation}
In this section, we evaluate the performance of the proposed  transmission techniques in terms of the achievable secrecy rate $R$ (bits/s/Hz). It is assumed that all communication takes place during a fixed coherence interval in which the channel  is assumed to known to both the transmitter and target receiver. Furthermore, for ease of exposition, we take a pessimistic approach and  assumed that the eavesdropper is aligned with one of the transmitter's $L$ paths.  

The secrecy rate $R$ is defined as
\begin{eqnarray}\label{Rs}
R = [\log_2(1+\text{SNR}_{\text{R}}) -\log_2(1+\text{SNR}_{\text{E}})]^+,
 \end{eqnarray}
where $\text{SNR}_{\text{R}}$ is the SNR at the target receiver, $\text{SNR}_{\text{E}}$  is the SNR at the eavesdropper, and $a^+$  denotes $\max\{ 0,a \}$. 
In the following we derive the average SNR expressions for the proposed transmission techniques.

\subsection{Random path selection}
From (\ref{yra88a}), the SNR at the receiver can be expressed as
\begin{eqnarray} \label{snrr1r}
\text{SNR}_{\text{R}} =  \frac{(\mathbb{E}[\alpha_i \sqrt{N/L} ])^2} {\sigma_\text{R}^2} =  \frac{N\bar{\alpha}^2} {L \sigma_\text{R}^2}=\frac{N\rho_\text{R}}{L},
\end{eqnarray}  
where $\bar{\alpha}$ is the average channel gain of all transmit paths, and $\rho_\text{R}=\frac{(\mathbb{E}[\alpha_i  ])^2} {\sigma_\text{R}^2} $. From (\ref{snrr1r}) we observe that the SNR at the target receiver deteriorates with increasing number of paths. This results as the total transmit power is spread among all $L$ paths and weaker links will be utilized for transmission.  However, as we will show in the derivation of the $\text{SNR}_{\text{E}}$ (see (\ref{snre1a})), increasing the number of paths increases the artificial noise at the eavesdropper. Hence there is an optimal number of paths $L$ that enhances the secrecy rate.

Similar to (\ref{snrr1r}), the SNR at an eavesdropper can be expressed as (see (\ref{arnl}))
\begin{eqnarray} \label{snre1a}\nonumber
\text{SNR}_{\text{E}}\hspace{-2mm} &=& \hspace{-2mm} \frac{1}{L}\bigg(\frac{\alpha_\text{E}^2 {N}} {L\sigma_\text{E}^2}\bigg) + \bigg(1-\frac{1}{L}\bigg) \bigg(\frac{\alpha_\text{E}^2(\mathbb{E}[ B(\theta_l) ])^2/LN} {\alpha_\text{E}^2 \text{var}[{B(\theta_l)}]/LN +\sigma_\text{E}^2}\bigg) \\
&=&  \frac{\rho_\text{E}N}{L^2}  + \bigg(1-\frac{1}{L}\bigg) \bigg(\frac{ \rho_\text{E} (\mathbb{E}[ B(\theta_l) ])^2} {\rho_\text{E} \text{var}[{B(\theta_l)}]+LN }\bigg)
\end{eqnarray}  
where $\rho_\text{E}=\frac{\alpha_\text{E}^2}{\sigma^2_\text{E}}$,  $\text{var}[\alpha_\text{E} B(\theta_l)/\sqrt{LN} ] =  \alpha_\text{E}^2/LN \text{var}[ B(\theta_l) ]$ represents the artificial noise variance, $\theta_l \in \Phi$, and 
\begin{eqnarray}
\mathbb{E}[ B(\theta_l) ] =  \frac{N}{L} \sum_{l=1}^L   \frac{\sin (N(\frac{\pi d}{\lambda} (\cos (\theta_\text{E}) - \cos (\theta_{l}))   )) } {\sin ((\frac{\pi d}{\lambda} (\cos (\theta_\text{E}) - \cos (\theta_{l}))   )) },
\end{eqnarray}  
and
\begin{eqnarray}
\text{var}[  B(\theta_l) ] =  \mathbb{E}[ B(\theta_l)^2 ] - (\mathbb{E}[ B(\theta_l) ])^2.
\end{eqnarray}  
The first term of (\ref{snre1a}) captures the eavesdropper SNR when it is aligned with one of the transmitter's AoD, i.e. $\theta_\text{E} = \theta_l$. This occurs with probability $1/L$ as it is assumed that $\theta_\text{E}$ overlaps with one of the receiver's AoDs, and there are a total of $L$ AoDs. The second term of (\ref{snre1a}) captures the eavesdropper SNR when it is not aligned with the transmitter's AoD, i.e. $\theta_\text{E} \not= \theta_l$. This occurs with probability $1-1/L$.  It is not difficult to observe from (\ref{snre1a}) that the eavesdropper's SNR deteriorates with increasing number of transmission paths.

\subsection{Joint path and antenna selection}
Using (\ref{yra8}), the SNR at the target receiver can be expressed as
\begin{eqnarray} \label{snrr1}
\text{SNR}_{\text{R}} &=&  \frac{(\mathbb{E}[\alpha_{\text{s}}M +\alpha_i (N-M)+\beta_\text{R} ])^2} {LN\sigma_\text{R}^2}\\
&=&   \frac{\alpha_{\text{s}}^2 M^2 } {LN\sigma_\text{R}^2} +   \frac{(\bar{\bar{\alpha}} (N-M)^2} {LN\sigma_\text{R}^2} +   \frac{(\mathbb{E}[\beta_\text{R} ])^2} {LN\sigma_\text{R}^2},
\end{eqnarray}  
where $\bar{\bar{\alpha}}$ is the average channel gain of all transmit paths, excluding the strongest path.

Similarly, using (\ref{ye1}) and (\ref{ye2}), the SNR at the eavesdropper can be expressed as
\begin{eqnarray} \label{snre1}\nonumber
\text{SNR}_{\text{E}} \hspace{-3.5mm} &=&  \hspace{-3.5mm}\frac{2}{L}\bigg(\frac{\alpha_\text{E}^2 (\mathbb{E}[{\hat{\beta}_\text{E}])^2}} {LN\sigma_\text{E}^2}\bigg) + \bigg(1-\frac{2}{L}\bigg) \bigg(\frac{\alpha_\text{E}^2(\mathbb{E}[ {\beta}_\text{E} ])^2} {\alpha_\text{E}^2 \text{var}[{{\beta}_\text{E}}] +LN\sigma_\text{E}^2}\bigg)\\
\hspace{-2mm}&=& \hspace{-2mm} \frac{2\rho_\text{E} (\mathbb{E}[{\hat{\beta}_\text{E}])^2}} {L^2N}+ \bigg(1-\frac{2}{L}\bigg) \bigg(\frac{\rho_\text{E}(\mathbb{E}[ {\beta}_\text{E} ])^2} {\rho_\text{E}\text{var}[{{\beta}_\text{E}}] +LN}\bigg),
\end{eqnarray}  
where the expectation is over all transmit antennas and AoDs. The first term of (\ref{snre1}) captures the eavesdropper SNR when it is aligned with one of the transmitter AoDs, i.e. $\theta_\text{E} = \theta_\text{s}$ or $\theta_\text{E} = \theta_{i}$, and  $\theta_{i},\theta_\text{s} \in \Phi $. This occurs with probability $2/L$. The second term of (\ref{snre1}) captures the eavesdropper SNR when it is not aligned with the transmitter's AoD, i.e. $\theta_\text{E} \not= \theta_i,\theta_\text{s}$. This occurs with probability $1-2/L$.  It is not difficult to see that $(\mathbb{E}[ \hat{\beta}_\text{E} ])^2<N^2$ in  (\ref{snre1}), and hence the first SNR term is lower than the first SNR term in (\ref{snre1a}).

\begin{figure}[t]
    \centering
    \includegraphics[width=250pt]{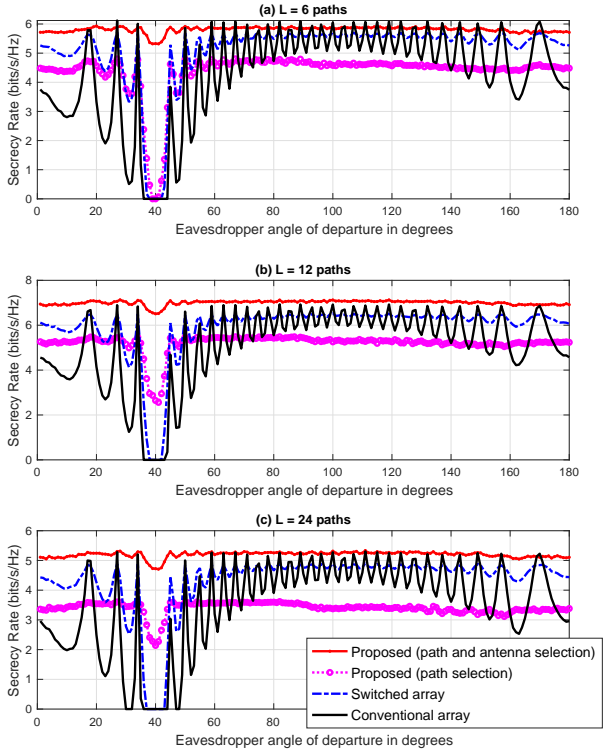} 
    \caption{Secrecy rate versus the eavesdropper's angle of departure $\theta_\text{E}$ for different number of transmission paths;  $N=32$ transmit antennas, $\rho_\text{R} = 10$ dB, and $\rho_\text{E} = 15$ dB.} \label{fig1}
  \end{figure}

\section{Numerical Results and Discussions}
In this section, we conduct numerical simulations to evaluate the efficacy  of the proposed techniques. We consider a setup where a transmitter is communicating with a receiver in the presence of an eavesdropper. Both the transmitter and receiver are unaware of the eavesdropper's location.
In this setup, the transmitter is equipped with a uniform linear array, with half wavelength separation, and the transmitter, receiver and eavesdropper have perfect knowledge of their channels. To benchmark the performance of the proposed techniques, we compare the secrecy rate achieved by the proposed techniques with the secrecy rate achieved when using conventional transmit antenna array, i.e. no action is taken at the transmitter, and the switched array technique proposed in \cite{sec5} since it can be applied on analog antenna architectures. For all setups, we assume that the strongest transmit path to the target receiver is along a fixed  AoD $\theta_\text{R} = 40$ degrees, i.e. $\theta_\text{R} = 40 \in \Phi$, and the remaining entries of $\Phi$ are randomly selected from the set $[1,180]$ degrees. The channel gains are assumed to be normally distributed with zero mean and unit variance. Both the conventional array  and switch antenna techniques always transmit along AoD $\theta_\text{R}  = 40$ degrees. When simulating the joint angle and antenna selection techniques, we assume that the secondary link is selected from the set of $L_\text{S} = 5$ strongest paths and $M=N/2$.

\begin{figure}[t]
    \centering
    \includegraphics[width=250pt]{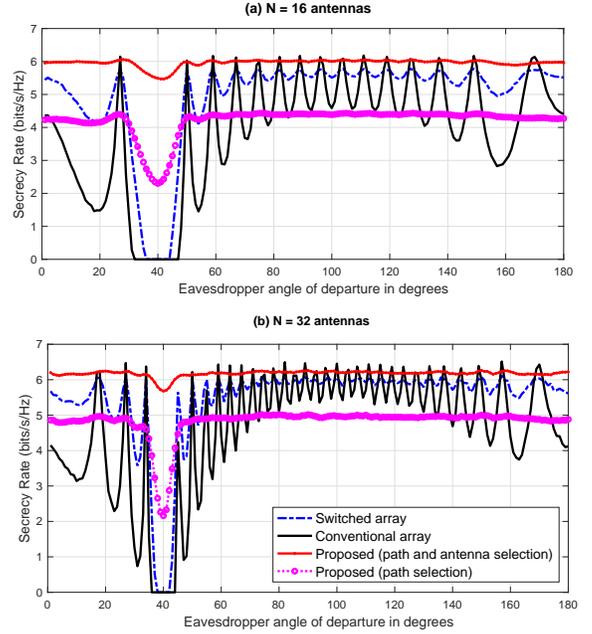} 
    \caption{Secrecy rate versus the eavesdropper's angle of departure $\theta_\text{E}$; $L=12$ transmit path,   $\rho_\text{R} = 10$ dB, and $\rho_\text{E} = 15$ dB.} \label{fig2}
  \end{figure}

To examine the performance of the proposed techniques, we plot the achievable secrecy rate when using the proposed techniques versus the eavesdropper AoD $\theta_\text{E}$ in Fig. \ref{fig1}. For all cases, we observe that the secrecy rate is higher when the eavesdropper AoD $\theta_\text{E}$ does not overlap with the transmission AoD $\theta_\text{R} = 40$ degrees. However, when  $\theta_\text{E} = 40$ degrees, i.e. the eavesdropper is located along one of the transmission path to the target receiver, we observe that all techniques suffer a performance hit, with the conventional technique and the switched array technique achieving zero secrecy rate for all values of $L$. The proposed path selection technique is shown to achieve higher secrecy rate for higher number of paths while the proposed path and antenna selection technique achieves the best performance for all cases. The reason for this is that for low number of paths, the probability of intercepting a direct link when using the path selection technique is higher. This reduces the secrecy rate. The proposed path and antenna selection technique achieves the best performance as it randomizes both the transmit angle and antennas associated with each angle, thereby resulting in higher artificial noise at the eavesdropper. 

In Fig. \ref{fig2}, we examine the effect of the number of transmit antennas on the secrecy rate. For all techniques, we observe that increasing the number of antennas increases the secrecy rate and the proposed techniques achieve non-zero secrecy rates at $\theta_\text{E} = 40$ degrees. This is achieved as  higher number of antennas result in  narrower beams and increased number of sidelobes. The increased number of sidelobe levels increases the variance of the artificial noise. The path selection technique is shown to achieve lower secrecy rate when the eavesdropper is not located along the main transmission path $\theta_\text{E} \not= 40$ degrees. This rate loss is experienced as the path selection technique does not always use the best transmission path, but rather, it transmits along a set of  paths, which contain weaker paths, to enhance the secrecy rate of the system.

\begin{figure}[t]
    \centering
    \includegraphics[width=250pt]{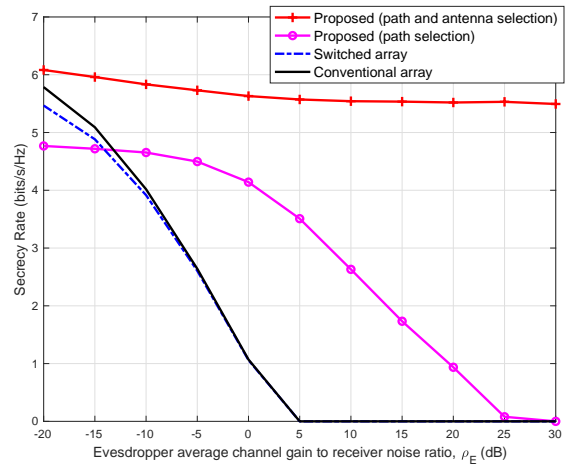} 
    \caption{Secrecy rate versus $\rho_\text{E}$;  
    $N=32$ transmit antennas, $\rho_\text{R} = 10$ dB, and $L= 12$ paths. Eavesdropper is located along AoD $\theta_\text{E} = 40$ degrees.} \label{fig3}
  \end{figure}

\begin{figure}[t]
    \centering
    \includegraphics[width=250pt]{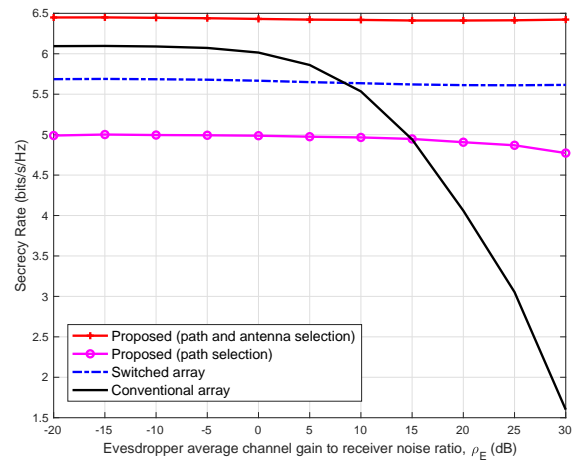} 
        \caption{Secrecy rate versus $\rho_\text{E}$;  
    $N=32$ transmit antennas, $\rho_\text{R} = 10$ dB, and $L= 12$ paths. Eavesdropper is located along AoD $\theta_\text{E} = 55$ degrees.}  \label{fig4}
  \end{figure}

To investigate the effect of the eavesdropper channel strength on the secrecy rate, we plot the secrecy rate versus the eavesdropper's average channel gain to noise ratio  $\rho_\text{E}$ when the eavesdropper is located along the transmitter's AoD $\theta_\text{E} = \theta_\text{R} = 40$ degrees in Fig. \ref{fig3}, and when the eavesdropper is located along the transmitters sidelobe, i.e. $\theta_\text{E} \not= \theta_\text{R}$ in Fig. \ref{fig4}. When the eavesdropper's communication link is weak, Fig. \ref{fig3} shows that all techniques achieve high secrecy rate. However, when the eavesdropper channel is stronger, the performance of both the switched-array and conventional array techniques plummet to zero, while the performance of the path selection technique deteriorates at a slower rate and the joint path and antenna selection technique deteriorates at a much slower rate as it achieves a higher artificial noise.  When the eavesdropper is not located along the transmitter's AoD, Fig. \ref{fig4} shows that all techniques, except the conventional array technique, achieve high secrecy rate. The proposed joint path and antenna selection technique achieves the highest secrecy rate as it randomizes both the transmit AoD and transmit antenna sets, followed by the switched-array technique, and the proposed path selection technique. The path selection technique achieves a lower secrecy rate as it uses all paths (with potentially weaker paths) for communications, while the path and antenna selection technique always uses the strongest path and an auxiliary path for transmission. This results in a higher secrecy rate.

\section{Conclusion}
This paper addressed the problem of PLS in the presence of an eavesdropper with overlapped channel paths with the target receiver. Two transmission techniques suitable for mmWave systems with analog antenna architectures are proposed. Both techniques use random transmission paths, instead of a single or all paths, for data transmission. This results in noise-like signals at an arbitrary eavesdropper and improves the secrecy of the communication system. Numerical results show that in the presence of scatters, the proposed techniques achieve significantly higher secrecy rate when compared to conventional and switched array techniques. Note that the proposed techniques require the number of paths $L>1$. For single path, i.e. LoS link, passive relays can be used to increase the number of transmission paths and induce randomness in the communication channel.

\section*{Acknowledgment}
This material is based upon work supported in part by the Sacramento State Research and Creative Activity Faculty Awards Program.


\end{document}